\documentstyle[graphicx,prb,aps]{revtex}
\def\f{$f$}
\begin{document}
\date{\today}
\title{
Correlation mechanism of the \f-electron delocalization.}

\author{U.\ Lundin, I.\ Sandalov, 
        O.\ Eriksson, and B.\ Johansson}

\address{Dept.\ of Physics, Uppsala University,
Box 530,SE-751 21 Uppsala, Sweden}  
\maketitle

\begin{abstract}
The mechanism of $f$-electron delocalization is 
investigated within the multi-orbital Anderson
lattice model by means of diagrammatic perturbation theory from the
atomic limit. The derived equations 
couple the intra-atomic transition energies, their spectral
weights and population numbers of the many-electron states. 
Its self-consistent solution for praseodymium metal shows that
the delocalization can be caused by external pressure via a resonant 
mixing of $f$- and conduction 
electrons in the vicinity of the the Fermi surface. It is also found that:
1. An increase of mixing leads to a decrease of the physical values of 
the Hubbard interactions, $U^*$, the reduction, however, is small. 2.
The initial Hubbard $U$ is split by renormalization into a set of
different physical values of $U^*_{i,j}$. 3. The gain in cohesive energy 
together with the $f$-sum rule 
cause a transfer of spectral weight, which is
decisive for the delocalization of \f-electrons. 
4. The correlated Fermionic quasi particles have their bandwidth
slightly reduced compared to the ones obtained by means of the 
Kohn-Sham equation.
\end{abstract}

\section{Introduction}
{\bf Published as: Phys.\ Rev.\ B {\bf 62}, 16370 (2000-II).}\\
The correct treatment of the local charge density built into the density 
functional method of {\it ab initio} band structure calculations, 
particularly in the local (spin) density approximation (LDA), provides a 
remarkably good description of those materials where the energy 
structure of the ions, established by the intra-atomic Coulomb interactions, 
is strongly modified by the band formation. If this condition is not 
fulfilled, like in many materials containing $d$- or \f-electrons, the 
band structure 
method is insufficient. The parameter which determines if the 
material belongs to this class of strongly correlated systems (SCES) is 
the ratio $W/U$, where $W$ is the bandwidth and $U$ is the parameter 
describing the Coulomb repulsion between $d$- or \f-electrons. There are a 
number of different suggestions of how to treat the failure of the LDA method. 
Also, at least three different definitions of the parameter $U$ 
in use. The first one was introduced by Hubbard~\cite{hubbI} as 
the Coulomb integral calculated between the wave functions of the 
corresponding $d$-, or \f-shells. Hubbard also discussed 
the importance of screening effects on $U$. This is mainly the Slater F$^o$ 
integral ($U_1$=F$^o$ is commonly used). 
The second definition, which comes from the formulation of the $s$-band 
Hubbard model, is $U_2$=[E$_{n+1}$-E$_n$]-[E$_n$-E$_{n-1}$], where E$_n$ is 
the energy of a $d$(\f)-atom with n electrons in the shell. An uncertainty 
exists also within this definition. The recipe used in the so-called 
LDA+U method~\cite{anisimovI} consists of calculation of the energies 
E$_{n-1}$ and E$_{n+1}$ of an atom either within an impurity-type or 
super cell approach. The results are then depending on where, and 
from where, the electron is removed, the degree of relaxation allowed for 
the charge on the 
neighboring atoms and the number of atoms allowed to be involved in the
relaxation. In the next step of the LDA+U method the found value $U_2$
is used for correcting the potential experienced by
$d$(\f)-electrons only, in spite of the fact that relaxation of charge
on the neighboring sites has been used when calculating $U_2$. 
Another approach is to use $U_2 \simeq \delta ^2E_{LDA}/\delta n^2_d $ as 
discussed by Gunnarsson {\it et al.\ } \cite{gunnarsson}.

It is clear from a physical point of view that 
the intensity of the correlations characterized by the value of $U$
should depend on the type of ground state and the approximation used.
Progress in this direction has been achieved
by Beiden {\it et al.\ } \cite{beiden,zein}. 
The definition used in their work,
$U_3=f^0_{dd}+f^0_{bb}-2f^0_{bd}$, 
 involves both the Coulomb matrix
elements in the linear muffin tin method using the spherical approximation 
(LMTO-ASA), $f^L_{ll'}$, and the average population numbers of different
shells, $f^0_{bb}=[f^0_{ss}n_s^2+f^0_{pp}n^2_p+2f^0_{sp}n_pn_s]/[n_p+n_s]^2$. 
Both the matrix elements and the population numbers, are
calculated self-consistently. The functional for these band structure 
calculations has been constructed on
the basis of the Gutzwiller-type approximation, and been applied to
 the 3-$d$ metals V,
Mn, Ni, Cu and 4-$d$ metals Nb and Tc. The Gutzwiller 
correlation correction influences the value of $U_3$ indirectly, via
the variational procedure for finding the population numbers. 

If we look at the definition of $U_2$, the physical values of $U^*_i$ are
determined by self-consistent values of the difference of the many electron
energies E$_n$. When we are discussing the ground state, or 
thermo-dynamical properties, all processes, fast and slow, renormalizing
E$_n$ have to be considered. 
The renormalization of $\Delta_{n,n-1}$=E$_n$-E$_{n-1}$ by long-range
Coulomb interaction~\cite{sandfil} leads to a dependence of these
energies on the shape of the Fermi surface via the frequency-dependent 
dielectric function. Peaks in the electron density of states 
amplify the effect of mixing and, therefore, cause a decrease of 
E$_n$-E$_{n-1}$, favoring a delocalization of $d$(\f)-electrons. 
In this paper we will consider renormalization of 
$\Delta_{n,n-1}$  caused by {\it kinematic} interactions due to 
mixing. This will be considered in the
present paper. The description of the high-energy experiments like
photo-electron spectroscopies, sometimes requires a different approach 
(see  Ref.\cite{johansson1}).

A very specific situation can arise in metallic \f-systems when one of the 
intra-atomic transitions happens to be in a close vicinity of the  Fermi 
level~\cite{kuzsand}. The scattering processes of the conduction electrons 
on this transition are resonantly amplified in this case. Therefore, one may 
expect a strong dependence of all renormalizing magnitudes on the mixing 
interaction if there is such a degeneracy.

In this paper we show that the delocalization of \f-electrons with pressure 
can be accounted for in a simple model. 
We also give quantitative numbers as regards the delocalization in Pr. 
Below, in order to display explicitly 
the mentioned effect of quasi-degeneracy 
we consider the periodical Anderson
model for the case of praseodymium metal in the 
corresponding region of parameters. 
Here, by means of a diagrammatic perturbation theory from the atomic limit,
we derive a system of self-consistent equations which determine the physical 
value of the Hubbard $U$, the spectral weights of the 
$[(n-1),n]$ and $[n,(n+1)]$ electron transitions. The equations include 
correlation effects explicitly, without
decoupling of the higher order correlation functions to the 
single-electron population numbers, and describe the dependence of the
physical properties involved 
on the degree of delocalization of the states. 
As will be seen below, this procedure always leads to a splitting of the
initial value for $U$ to many different orbital-dependent $U^*$. Here, we 
only inspect the
orbitally polarized solution, the effect of crystal-field levels is not 
taken into account, since the scale of the crystal-field transitions is
much smaller than the ones considered here, although it can be included
in the formalism.

\section{The Model and its approximation}

We start with the periodical multi orbital Anderson Hamiltonian,
\begin{equation}
{\cal H}=\sum_{{\bf k}, \mu}
\epsilon^{\vphantom{\dag}}_{{\bf k} \mu}c^{\dag}_{{\bf k} \mu}
c^{\vphantom{\dag}}_{{\bf k}\mu}
+\sum_{i,\mu} \epsilon^0 \hat{n}_{i\mu}  
+ {1 \over 2}\sum_{i,\nu
\neq \mu} U_{\nu \mu} \hat{n}_{i\nu}  \hat{n}_{i\mu}  
+ \sum_{\mu,{\bf k},i}[ V_{\mu}({\bf k})e^{i{\bf k
  R}_i}c^{\dag}_{{\bf k} \mu}f^{\vphantom{\dag}}_{i \mu}+ 
V^*_{\mu}({\bf k})e^{-i{\bf k
R}_i}f_{i \mu}^{\dag}c^{\vphantom{\dag}}_{{\bf k} \mu}],
\end{equation}
where
$\hat{n}_{i\mu} = f^{\dag}_{i\mu} f_{i\mu}$, $i$ is the site index and
$\mu$ is the \f-orbital index, i.e. the $\sigma,m_l$ states.
The first term describes the non-\f conduction electrons, 
$\epsilon_{{\bf k} \mu}$ is the band structure energy, which in our 
case is taken 
from a self consistent band structure calculation. The two next terms 
describe the strongly correlated \f-electrons, and the last term represents 
the mixing interaction between the \f-states and the conduction band.
Thus, we will study self-consistently the changes in the band structure 
caused by the mixing interaction in the background of the strongly correlated 
\f-system. 
We define many-particle \f-states, $\Gamma$, as
\begin{eqnarray*}
&&|\Gamma_0 \rangle  \stackrel{{\rm def}}{=}| 0 \rangle \\
&&|\Gamma_\mu \rangle  \stackrel{{\rm def}}{=}|\mu \rangle = 
f^{\dag}_\mu|0 \rangle \\
&&|\Gamma_{\mu,\mu'} \rangle \stackrel{{\rm def}}{=}|\mu\mu' \rangle = 
f^{\dag}_\mu f^{\dag}_{\mu'}|0 \rangle, \cdots 
\end{eqnarray*}
and so on for all 14 \f-orbitals.

The creation and annihilation operators for the \f-electrons are
rewritten in terms of Hubbard $X$-operators, describing 
transitions between many-particle states $\Gamma$ and $\Gamma'$, 
$X_i^{\Gamma,\Gamma'}\stackrel{{\rm def}}{=}|\Gamma\rangle\langle \Gamma'|$. 
The expansion is written as 
\begin{equation}
\label{expansion}
f_{i\mu} = \sum_{\Gamma,\Gamma'} \langle\Gamma|f_{i\mu}|\Gamma'\rangle 
X_i^{\Gamma',\Gamma} =  \sum_{\Gamma}\langle\Gamma|f_{i\mu}|
\mu\Gamma\rangle X_i^ {\mu\Gamma,\Gamma}.
\end{equation}
We define the transitions $\Gamma$ to $\mu\Gamma$ ($\Gamma$ to $\Gamma'$) 
as $a$, and the inverse, $\mu\Gamma$ to $\Gamma$, as $\bar{a}$.
Now let us reformulate the Hamiltonian and  the
Green functions (GF's) for the \f-operators, in terms of Hubbard operators 
and then use a diagram technique for the $X$ operators~\cite{izyum}.
The electronic Matsubara temperature \f-Green functions should be expressed 
in terms of $X$-operator GF's 
\begin{equation}
\langle{\cal T} f^{\vphantom{\dag}}_{i\mu}(\tau) 
f^{\dag}_{i\mu'}(\tau')\rangle 
  =\sum_{\Gamma_1,\Gamma_2}  
   \langle\Gamma_1|f_{i\mu}|\mu\Gamma_1\rangle 
   \langle\mu'\Gamma_2|f^{\dag}_{i\mu'}|\Gamma_2\rangle *
   \langle{\cal T} X_i^{\Gamma_1,\mu\Gamma_1}(\tau)
    X_i^{\mu'\Gamma_2,\Gamma_2}(\tau')\rangle.
\end{equation}
The $X$-operators satisfy the commutation relation
\begin{equation}
\label{commutator}
\left[ X^{p,q}_i , X^{r,s}_j \right]_\eta = \delta_{ij} \left( \delta_{qr}
X^{p,s} + \eta \delta_{sp} X^{r,q} \right).
\end{equation}
The fact that the (anti)commutator of two $X$ operators is still
an operator gives rise to ''end'' factors (averages 
of one $X$-operator, $\langle X^{\Gamma,\Gamma}\rangle$) when the extended 
Wick's theorem is
applied to the Green functions. In the diagram technique end factors are
represented by open (closed) circles for the zero (dressed) spectral
weights. The zero Green functions for $X$, and $c$-operators are 
\begin{eqnarray}
&&\left(G_{m,n}^{a,\bar{b}}(i\omega_n)\right)^0  \stackrel{{\rm def}}{=}  -i
\langle{\cal T} X_n^a(\tau)X_m^{\bar{b}}(\tau')\rangle^0_\omega  
= \delta(a-b) \frac{P_a^0}{i\omega_n -\Delta_a^0}\delta_{m,n} 
\stackrel{{\rm def}}{=} P_a^0{\cal D}_a^0,\\
&&\left(C_\mu({\bf k},i\omega_n)\right)^0 \stackrel{{\rm def}}{=} 
-i\langle{\cal T} c_{{\bf
k}\mu}(\tau)c^{\dag}_{{\bf k} \mu}(\tau')\rangle^0_\omega = 
\frac{1}{i\omega_n-\epsilon_{{\bf k}\mu}} ,
\end{eqnarray}
where $i\omega_n=(2n+1)\pi T$ and $T$ is the temperature in energy units 
(k$_B$=1). $P_a$ denotes the end factors, 
$P_a=P_{\Gamma,\Gamma'}=\langle \{ X^{\Gamma,\Gamma'},
X^{\Gamma',\Gamma}\}\rangle=
\langle X^{\Gamma,\Gamma}\rangle+\langle X^{\Gamma',\Gamma'}
\rangle=N_\Gamma+N_{\Gamma'}$,  
which in lowest order are spectral weights, and 
$\Delta_{\Gamma',\Gamma}=E_{N_{\Gamma'}}-
E_{N_{\Gamma}}$ 
the energy of the transition $a$ entering the locators~\cite{sanda}, ${\cal
D}_a(i\omega_n)=\frac{1}{i\omega_n-\Delta_a}$, for the 
$X$-operator Green functions. Averaging the expansion 
(\ref{expansion},\ref{commutator}) for the anticommutator
\begin{equation}
1=\{ f_{i,\mu},f^{\dag}_{i,\mu} \} = 
\left \{ \sum_{\Gamma}X_i^{\Gamma,\mu\Gamma},\sum_{\Gamma'}
X_i^{\mu\Gamma',\Gamma'} \right \} 
= \sum_{\Gamma,\Gamma'} \{X_i^{\Gamma,\mu\Gamma},X_i^{\mu\Gamma',\Gamma'}
\},
\end{equation}
gives a sum rule for the spectral weights in the $X$-operator Green
functions.

\begin{center}
\begin{figure}
\includegraphics[scale=0.7]{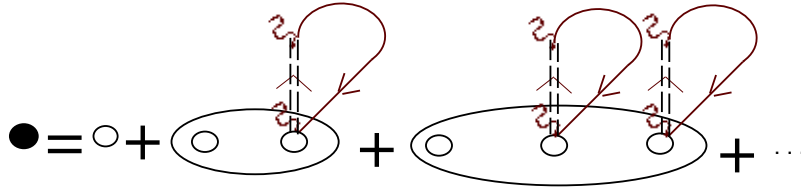}
\caption{ The correction $\delta P_a^{(1)}$ to the population numbers.}
\label{pop2:graph}
\end{figure}
\end{center}

Using a diagram technique for the Hubbard operators~\cite{izyum} we formulate 
a mean field theory (defined as the theory with no energy dependent
corrections to the self energy, $\Sigma$, and to the spectral weights 
$P_{\Gamma,\Gamma'}$). The second order graphs are 
summed to give corrections to the population numbers, $P_a=P_a^0+\delta
P_a^{(1)}+\delta P_a^{(2)}$, and the frequencies,
$\Delta_a=\Delta_a^0+\delta \Delta_a$. Within the Fermionic hierarchy, 
the first correction to the population numbers, 
the Ising-model-like, $\delta P_{a}^{(1)}$, is given by the sum of 
ovals, and is shown in Fig.\ref{pop2:graph}. 
We denote the conduction
Green functions by dashed lines, the locators for $X$-operator Green 
functions by full lines, 
bold full lines and double dashed lines mean the full Green functions, and
the dots denote the terminal factors (local spectral weights). 

The cumulant contribution, $\delta P_{a}^{(1)}$ (see Fig.\ref{pop2:graph}) 
can be calculated from a 
generating functional as described by Sandalov {\it et al.\ }~\cite{sanda}.
\begin{center}
\begin{figure}
\includegraphics[scale=0.8]{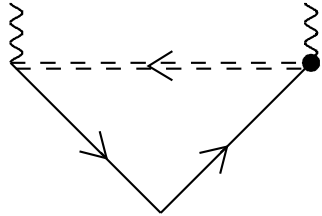}
\caption{The correction $\delta P_a^{(2)}$ to the population numbers.}
\label{pop:graph}
\end{figure}
\end{center}
This procedure gives the expansion of an exponential function. 
However, this contribution, in the paramagnetic state, does not change the 
relative weights of the population numbers and can be omitted.
The second contribution $\delta P_{a}^{(2)}$, see Fig.\ref{pop:graph}, 
is given by 
\begin{equation}
\label{pop:corr}
-\beta^{-1}\sum_{\omega_n,\mu,a_{1},a_{2}}
T_{\mu}(i\omega_n) f^{a_{1}}_{\mu}(f^\dagger_{\mu})^{\bar{a}_{2}}
{\cal D}_{a_{1}}(i\omega_n) \kappa_{a_{2}}^{a_{1};a,\bar{a}} 
{\cal D}_{a_{2}} (i\omega_n)P_{a_{2}},
\end{equation}
where $\kappa$ are structure constants of the algebra and
$\beta^{-1}=T$ is the temperature and 
\begin{equation}
T_{\mu}(i\omega_n)=\sum_{{\bf k},\mu}V^*_{\mu}({\bf k})
C_{{\bf k},\mu}(i\omega_n) V_{\mu}({\bf k}).
\end{equation}

\begin{center}
\begin{figure}
\includegraphics[scale=0.8]{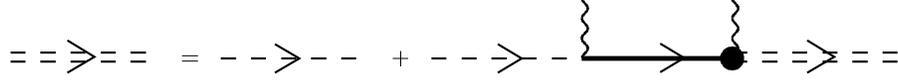}
\caption{The graph for the full Green function for the conduction electrons.}
\label{cond:graph}
\end{figure}
\end{center}

\begin{center}
\begin{figure}
\includegraphics[scale=0.8]{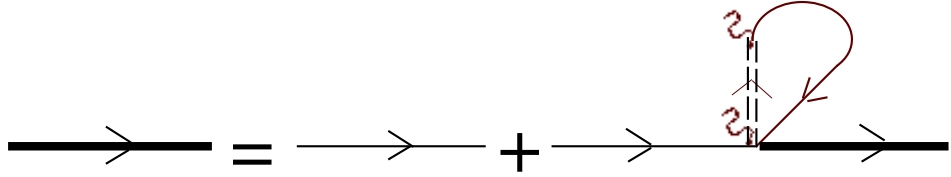}
\caption{Correction to the frequencies $\delta\Delta_a$.}
\label{freq:graph}
\end{figure}
\end{center}

The Green function for the conduction electrons are dressed by 
the interaction as shown in Fig.\ref{cond:graph} 
to give 
\begin{equation}
\label{conduct}
\left[ C_{\mu}({\bf k}, i \omega_n)\right] ^{-1} =
\left[ C_{\mu}^{(0)}({\bf k}, i \omega_n)\right] ^{-1} - \sum_a
|V_{\mu}({\bf k})|^2|f_{\mu}^{a}|^2
P_a {\cal D}_a(i \omega_n).
\end{equation}
The shift of the frequencies, 
$\delta\Delta_{a}=\Delta_{a}-\Delta_{a}^{0}$, comes from the equation 
shown in Fig.\ref{freq:graph}, 
and is given by
\begin{equation}
\label{freq:corr}
-\beta^{-1}\sum_{\omega_n,\mu} T_{\mu}(i\omega_n)
{f^{a}_{\mu}}(f^{\dagger}_{\mu})^{\bar{a}}{\cal D}_{a}(i\omega_n).
\end{equation}
These equations are coupled and accordingly have to be solved 
self consistently. Note 
that, so far, this formalism is not restricted to the case of praseodymium, 
it is valid for any \f-electron system.
The physics of \f-systems is described by the positions of the levels 
$\Delta_a$ and the corresponding spectral weights $P_a$. 

\section{Application to praseodymium}

\begin{center}
\begin{figure}
\includegraphics[scale=1.0]{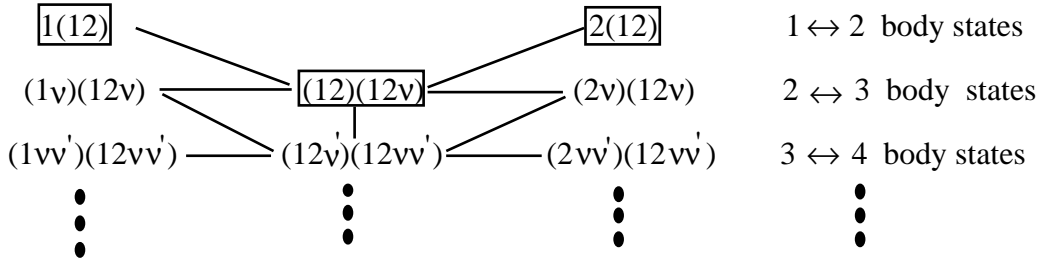}
\caption{Possible transitions. The boxes indicate
the states which have nonzero spectral weight in the start
configuration.}
\label{trans}
\end{figure}
\end{center}

We now concentrate on the model for praseodymium metal. 
The existence of crystal-field effects says that Pr is in an orbitally
polarized $f^2$ state.  Two
localized $f$-electrons are placed into  orbitals 1 and 2;
this two-electron state is denoted as $|$12) (in indices as (12)).
When mixing is absent $n_f =2$ and $N^0_{(12)} =1$. Switching the
 mixing on causes transitions to $f^1$ and $f^3$ states; we denote
$f^1$ as 1 or 2 depending on $m_l$, $f^3$ as $|12\nu)$, $\nu$ being one
of the 12 unoccupied $f$-orbitals. The way that spectral weight transfers 
to higher transitions is shown in Fig.\ref{trans}. 
The energies for these states are
estimated from $E_{n}=n\epsilon^0 + Un(n-1)/2$ with $\epsilon^0$ providing
the energy minimum for $n=2$ and describing the attraction to the nucleus. 
This gives $\epsilon^0=-3U/2$. 
The bare energies of the transitions are, 
from $|$12) to $|$1) or $|$2), denoted by indices as (12)1 and (12)2,
$\Delta_1 = E_2-E_1 = \Delta_{(12)1} = \Delta_{(12)2} = -U/2$,
 from $|$12$\nu$) to $|$12)  $\Delta_2 = E_3-E_2 = 
\Delta_{(12\nu)(12)} = U/2=\Delta_{(12\nu)(1\nu)}=\Delta_{(12\nu)(2\nu)}$, 
and from $|$12$\nu\nu'$) to $|$12$\nu$) $\Delta_{3} = 3 U/2 $.
The zero \f-Green functions can be found by inversion of the 
expansion in Eqn.~\ref{expansion}, 
$F_{1}^{(0)}=\langle T f_1 f_1^{\dagger}\rangle^0 = F_{2}^{(0)}$ and
$F_{\nu}^{(0)}=\langle T f_\nu f_\nu^{\dagger}\rangle^0$ becomes 
\begin{equation}
F_{1}^{(0)}=\frac{P_{(12)2}^0}{i\omega_n -\Delta^0_{(12)2}} +
\sum_\nu \frac{P_{(12\nu)(2\nu)}^0}{i\omega_n -\Delta^0_{(12\nu)(2\nu)}} 
+\cdots = \frac{1}
{i\omega_n -\Delta^0_{(12)2}} ,
\end{equation}
using the expansion (Eqn.\ref{expansion}), and for the 12 different 
$\nu$ orbitals
\begin{equation}
F_{\nu}^{(0)}=\frac{P_{(12\nu)(12)}^0}{i\omega_n -\Delta^0_{(12\nu)(12)}} 
 + \sum_{\nu'} (1-\delta_{\nu\nu'})\frac{P_{(12\nu\nu')(12\nu)}^0}
{i\omega_n -\Delta^0_{(12\nu\nu')(12\nu')}} +\cdots =
\frac{1}{i\omega_n-\Delta^0_{(12\nu)(12)}}.
\end{equation}
Then, the dressed conduction Green functions (\ref{conduct}) can be written
as,
\begin{equation}
C_{\mu}({\bf k}, i \omega_n)=\frac{1}{i \omega_n - \epsilon_{\bf k}^{\mu}-
|V_{\mu}({\bf k})|^2 F_{\mu}(i\omega_n)} ; \mu =1,2,\nu.
\end{equation}

Due to the complex structure of the system of equations we make some further
simplifications. We do not allow non-diagonal hopping 
(this case has been considered in ref.~\cite{md}) to occur.
Non-diagonal terms affect the shifts only in fourth order of perturbation 
theory and are thus much smaller than the effects considered here. Physically 
the non-diagonal hopping would give a small width to the levels, not 
changing the physical picture to a large extent. 
Further we assume that the mixing parameter, $V_{\mu}({\bf k})$, has a {\bf k}
dependence following $\epsilon_{\bf k}$. Then the summation over {\bf k}
in the formulas above can be replaced by a integration over the density
of states (DOS) $g_{\mu}(\epsilon)=\sum_{\bf k}\delta(\epsilon-\epsilon_{\mu
{\bf k}})$. 
Applying the formula (Eqn.~\ref{freq:corr}) for the transition $a$=(12)1 
we obtain
\begin{equation}
\delta\Delta_{(12)1}=\sum_{\omega_n,{\bf k}} |V_2({\bf k})|^2
{\cal D}_{(12)1}(i\omega_n) C_2({\bf k},i\omega_n)
\end{equation}
and transforming summation over the Matsubara frequencies to a integration
over energy, we obtain an expression in terms of the retarded Green functions 
\begin{equation}
\label{freq:Pr}
\delta\Delta_{(12)1}=\int_{-\infty}^{\infty}d\omega f(\omega-\mu)
\int_{W_a}^{W_b} dx
g_2(x)|V_2(x)|^2 
\left( -\frac{1}{\pi} \right) Im \left[
\frac{1}{\omega+i\delta-\Delta_{(12)1}}\cdot C_2^R(x,\omega+i\delta) \right
]
\end{equation}
The correction to the spectral weight according to Eqn.(~\ref{pop:corr}) 
becomes
\begin{eqnarray}
\label{pop:Pr}
&&\delta P_{(12)1}=\int_{-\infty}^{\infty}d\omega f(\omega-\mu)\int_{W_a}^{W_b} 
dx g_2(x)|V_2(x)|^2 
\left( -\frac{1}{\pi} \right) Im \left[
\frac{P_{(12)2}}{\left[\omega+i\delta-\Delta_{(12)2}\right]^2}\cdot
C_2^R(x,\omega+i\delta) \right ]  \nonumber \\
&&- \sum_{\nu} \int_{-\infty}^{\infty}d\omega f(\omega-\mu)\int_{W_a}^{W_b} 
dx g_{\nu}(x)|V_{\nu}(x)|^2 
\left( -\frac{1}{\pi} \right) Im \left[
\frac{P_{(12\nu)12}}{\left[\omega+i\delta-\Delta_{(12\nu)(12)}\right]^2}
\cdot C_{\nu}^R(x,\omega+i\delta) \right ].
\end{eqnarray}
Note that the contributions from the lower and upper Hubbard sub-bands
enter with different signs, i.e.\ these contributions favor 
a localization and a delocalization respectively. 
Due to symmetry of the corrections the (12)1 transition gives the 
same corrections as
(12)2, and (12$\nu$)(1$\nu$) the same as (12$\nu$)(2$\nu$). 
The corrections which we have taken into account are listed in tables
\ref{tab:freq} and  \ref{tab:pop}.
\begin{table}[htb]
\begin{tabular}{c|l}
$a$ & $\delta\Delta_a$ \\ \hline
(12)(1)           & $\langle                     |V_2|^2 {\cal
                    D}_{(12)1}(i\omega_n)
                    C_2 \rangle_{\bf k}$   \\
(12$\nu$)(1$\nu$) & $\langle                     |V_2|^2 {\cal
                    D}_{(12\nu)(1\nu)}(i\omega_n)
                    C_2 \rangle_{\bf k}$   \\
(12$\nu$)(12)     & $\langle                     |V_\nu|^2  {\cal
                    D}_{(12\nu)(12)}(i\omega_n)
                    C_\nu \rangle_{\bf k}$   \\
(12$\nu\nu$')(12$\nu$') &$11\langle              |V_\nu|^2
                     {\cal D}_{(12\nu'\nu)(12\nu')}(i\omega_n)
                          C_\nu \rangle_{\bf k}$  \\
\end{tabular}
\caption{Corrections to the frequencies. Note that only 11 of the 12
unoccupied $\nu$ orbitals (i.e. the $\nu'$ orbitals) contribute to the
shift
of the $f^4-f^3$ transition.
(Summation over k and integration over frequency is denoted as,
$\sum_{\bf k}\cdots \stackrel{{\rm def}}{=}\langle\cdots\rangle_{\bf
k}$).
}
\label{tab:freq}
\end{table}
\begin{table}[h]
\begin{tabular}{c|l}
$a$  &  $\delta P_a$  \\ \hline
$(12)1$ &
  $\begin{array}{l} \langle |V_1|^2
[{\cal D}_{(12)2}
   ]^2
  P_{(12)2} C_1 \rangle_{\bf k} - \\
  - 12\langle
   |V_\nu|^2
  [{\cal D}_{(12\nu)(12)}]^2
  P_{(12\nu)(12)} C_\nu \rangle_{{\bf k},\nu}
\end{array}$
                                                \\
 & \\
$(12\nu)(1\nu)$ &
  $\begin{array}{l} \langle |V_2|^2
[{\cal D}_{(12\nu)(2\nu)}
   ]^2
  P_{(12\nu)(2\nu)} C_2 \rangle_{\bf k} - \\
  - \langle
   |V_\nu|^2
  [{\cal D}_{(12\nu)(12)}]^2
  P_{(12\nu)(12)} C_\nu \rangle_{\bf k}
\end{array}$
                                                \\
 & \\
$(12\nu)(12)$     &
  $\begin{array}{l} \langle                |V_2|^2
  [{\cal D}_{(12)1}]^2 P_{(12)1} C_2 \rangle_{\bf k} + \\
  + \langle                |V_1|^2
  [{\cal D}_{(12)2}]^2 P_{(12)2} C_1 \rangle_{\bf k} \end{array}$
                                                \\
 & \\
$(12\nu\nu')(12\nu')$ &
  $\begin{array}{l}11\langle               |V_\nu|^2
   [{\cal D}_{(12\nu\nu')(12\nu')}]^2
     P_{(12\nu\nu')(12\nu')}C_\nu \rangle_{\bf k} \end{array}$
                                                      \\
\end{tabular}
\caption{Corrections to the spectral weights, the factor 11 comes from
11 of the 12 unoccupied $\nu$ orbitals (i.e. the $\nu'$ orbitals).}
\label{tab:pop}
\end{table}

The bare band structure of Pr metal consists of conduction bands, a local
level, $\Delta_1$, below the bottom of the conduction band,
a level $\Delta_2$ slightly above the
Fermi energy, $\epsilon_F$, and $\Delta_3$, which
is much higher than $\epsilon_F$. The
input DOS is shown in Fig.\ref{result2} (V=0).
The transitions, $\Delta_{1}^{\pm1}$, from the crystal-field levels, 
$|\pm 1\rangle$, 
are slightly above the transition $\Delta_{1}$ of the singlet state and
are empty in the ground state. We do not write them here explicitly, since the
summation over the CF-levels gives the same for the lower spectral weight. 
Therefore, at T=0 K, this does not change the delocalization scenario.

\section{Analysis and Conclusions}
The equations (\ref{pop:corr})-(\ref{freq:corr}) represent the general
form of the mean-field equations for the multi-orbital Anderson model 
and some conclusions of
general nature can be drawn from them. 
{\it First}, as seen in the type of solution considered above, contrary to the
Gutzwiller~\cite{gutzwiller} approach~\cite{beiden,zein}, 
the description of excitations in terms of \f-orbitals becomes meaningless 
since many-Fermion magnitudes are involved
and the proper way to describe electrons is in terms of
collective atomic like excitations described by the Hubbard operators. 
The Fermi-like \f-excitations  are
separated into different subgroups. The part described by the lower
poles of the 
$f$-electron Green function (Fermionic transitions), is not important 
for thermodynamics.
Inspection of the next corrections to these spectral weights shows
that they depend on the spin dynamics. The upper-pole excitations form
delocalized states, contributing to the cohesive energy, 
and the corresponding band structure depends mainly
on the $f$-spectral weights in the upper pole. The latter
being coupled by the sum rule to the lower pole, reflect the dynamics of
localized spins and, therefore, includes in principle temperature effects 
into the band structure in a different way compared to usual approaches.
{\it Second}, the upper and lower frequencies ($\Delta_2$ and
$\Delta_1$) are moved due to
renormalization by mixing interaction towards each other,
giving a reduction of the effective $U^*$. All other frequencies are
also renormalized, thus leading to a splitting of the 
Hubbard $U$ into several band structure dependent $U_{a,a'}^*=
\Delta_a^{(n+1,n)}-\Delta_{a'}^{(n,n-1)}$. 
{\it Third}, when
mixing is turned on, part of the electrons are placed into the mixed state
$\langle c^\dag_{{\bf k}\mu}X_i^{\Gamma,\Gamma'}\rangle$, and the sum rule
for the total spectral weight causes the {\it occupation of the lower pole to
decrease} with the same amount, i.e.\ the system  undergoes delocalization.  
Therefore, in the considered solution {\it the
value of the observable local moment turns out to be linked to the strength of
the mixing interaction in the
other energy region, near} $\epsilon_F$. Particularly, the localized
moment observed in neutron
experiments and the Curie constant in high temperature magnetic susceptibility
should be proportional to the spectral weight of the lower
pole. Note that this mechanism is difficult to extract from the 
one-electron GF's 
since the transfer of the spectral weight is hidden in the complex 
many-electron 
vertices, and cannot be obtained in a one-electron picture at all. 
This mechanism of delocalization does not work if the
mixing interaction is self-consistently switched off by the correlations.

The correction $\delta P^{(2)}$ to the spectral weights contain the locator
${\cal D}_a(i\omega)$ in the second power, whereas the frequency shift 
$\delta\Delta_a$ only in the first. Therefore, $\delta P^{(2)}$ depends on the
critical parameter $\Delta_{\Gamma\Gamma'}-\mu$ much stronger than the shift
$\delta\Delta_a$ and it is the spectral weight transfer, not the decrease of
the Hubbard $U$, which 
is decisive for the delocalization of \f-electrons
under applied pressure for the scenario considered. 
{\it Increasing the mixing, under external pressure, causes 
spectral weight to transfer from the $\Delta_1$- to the 
$\Delta_2$-energy region.}

Within this model, where hopping is not taken into account, 
the mixing of ($spd$)-states with the transition $\Delta_2$ forms a pseudo-gap
near $\Delta_2$, the DOS increases in the vicinity of
$\epsilon_F$  and
electrons fill in additional states, therefore the Fermi energy 
$\epsilon_F$  moves down. Thus,
{\it part} of $f$-{\it Hubbard's transitions } influence the
cohesive energy of the metal. Composing 
the $f$-electron's Green functions  spectral weights from the $X$-operator 
Green functions, and taking small
enough values of the mixing interaction, we find that 
the same statement is valid also for $f$-{\it electrons}.
Thus, the energy needed for this transfer is provided by the $PV$ term in the
free energy and by the contribution from \f-electrons to the cohesive energy. 

At last, some conclusions specific to $Pr$ can be added to what's given 
above. The deformation of the band structure also results in an 
increase of the effective mass, which should 
give an enhancement of the Pauli-part of the magnetic susceptibility 
besides the crystal-field mechanism described by White and Fulde~\cite{white}.

However, as described above, the value of the localized $f$-moment should be 
slightly reduced. Indeed, switching on mixing leads to 
the shifts: $N^0_{(12)}\rightarrow N^0_{(12)}-\delta N^0_{(12)}$, 
$N_{(12\nu)}^{(0)}=0\rightarrow$ $N_{(12\nu)} \sim n_{cX} 
\stackrel{{\rm def}}{=}\sum\langle c^{\dag}_{k\nu} 
X^{(12)(12\nu)}_{k}\rangle\neq 0$. This increases 
the spectral weight in the upper Hubbard band to $12n_{cX}$. 
Since the total spectral weight of $G_{(12)}^f$ is equal to one, 
the spectral weight in 
a lower pole, at $\omega \sim \Delta_{(12)1}$, 
decreases by the same amount. 

\section{Numerical Calculations for praseodymium}
We made two numerical tests, one with a square shaped DOS using a mixing in 
the form $V_{\bf k}=V_1+V_2(\epsilon_{\bf k}-\epsilon_0)$ and varying 
the constants $V_1$ and $V_2$ and the position of the levels. 
The second type of numerical test were 
done for a DOS for the conduction band of the Pr metal. The DOS was 
obtained using a full-potential LMTO method 
with the two 4\f-electrons in the core and for the real dhcp structure 
(see Fig.\ref{result2} V=0). In this case we used $V_{\bf k}=V_1$ only. 
According to Anderson's argument~\cite{anderson} due to symmetry 
reasons the ${\bf k}$-independent part of the single-electron mixing should 
vanish. There is, however, a local contribution to mixing from Coulomb 
interaction~\cite{sandalov}, 
\begin{equation}
\bar{V}_1c^\dag_{\vphantom{1}} f^\dag_{\vphantom{1}} f f= \bar{V}_1 
\langle \gamma |f^\dag_1f^{\vphantom{\dag}}_2f^{\vphantom{\dag}}_3|
\Gamma\rangle 
c^\dag_{\vphantom{1}} X^{\gamma,\Gamma}_{\vphantom{1}}=
Vc^\dag_{\vphantom{1}} X^{\gamma,\Gamma}_{\vphantom{1}}, 
\end{equation}
which is allowed to be non-zero. 
The non-local part we model by the term $V_2(\epsilon_{\bf k}-\epsilon_0)$ 
providing $\sum_{{\bf k}}V_2(\epsilon_{\bf k}-\epsilon_0)=0$. Thus, we assume 
that the main contribution to the ${\bf k}$-dispersion comes from the 
neighboring atoms which gives the ${\bf k}$-dependence in both 
$\epsilon_{\bf k}$ and $V_{\bf k}$. We choose $U$ to be 10 eV.
The system of equations was solved by means of a 
steepest descent method. The criterion for 
convergence used was that the sum of the squared difference
of the input and output between two
consecutive iterations, i.e.\ $\sum_a (\delta P_a - \delta
P_a^{(calc.)})^2 + \sum_a(\delta \Delta_a - \delta
\Delta_a^{(calc.)})^2$ should be less than $10^{-6}$.
It is worth to mention that it is very difficult to achieve convergence near 
the transition from localized to delocalized and that we had to mix a very 
small portion of a previous iteration into the new solution. 
Further, a good start guess was important for the convergence. 
\subsection{Constant DOS Case}
\begin{center}
\begin{figure}[hbt]
\includegraphics[scale=1.0]{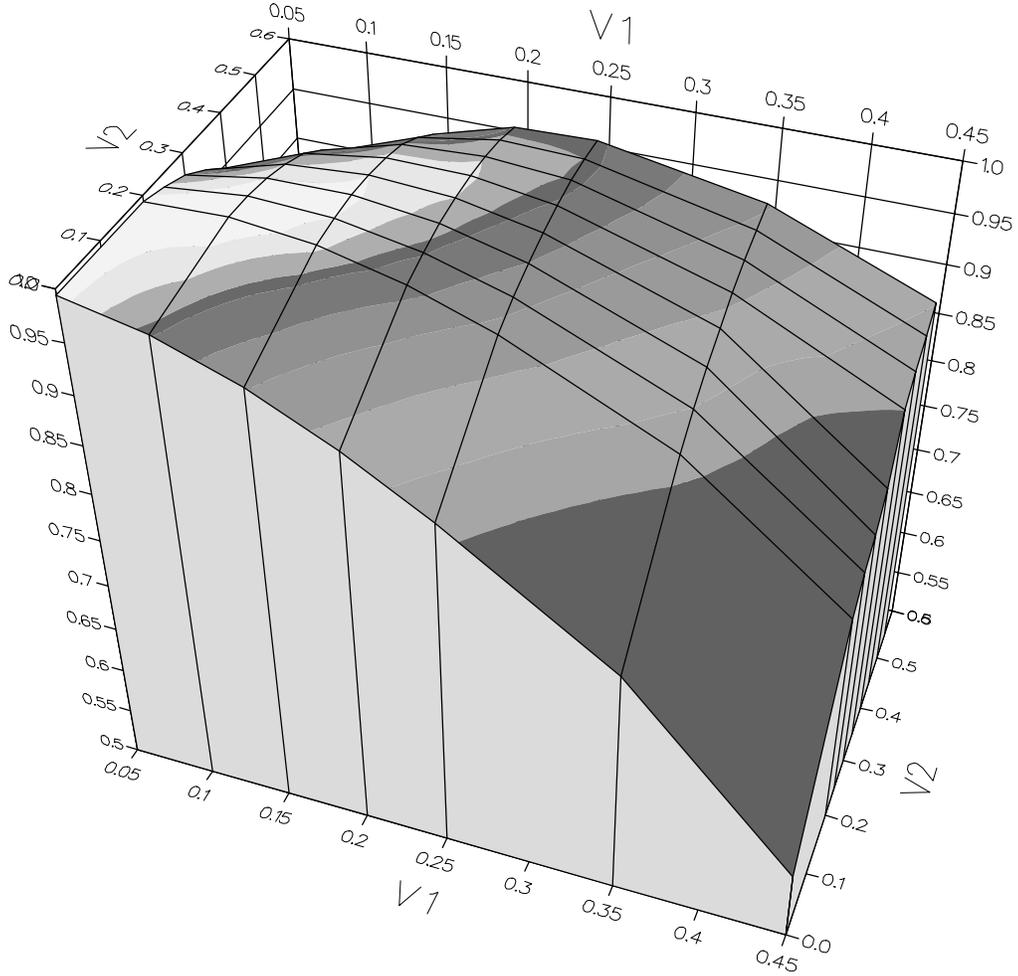}
\caption{Population in the lower pole, when changing the parameters $V_1$ and
$V_2$.}
\label{pv1pv2:fig}
\end{figure}
\end{center}

\begin{center}
\begin{figure}[hbt]
\includegraphics[angle=270,scale=0.6]{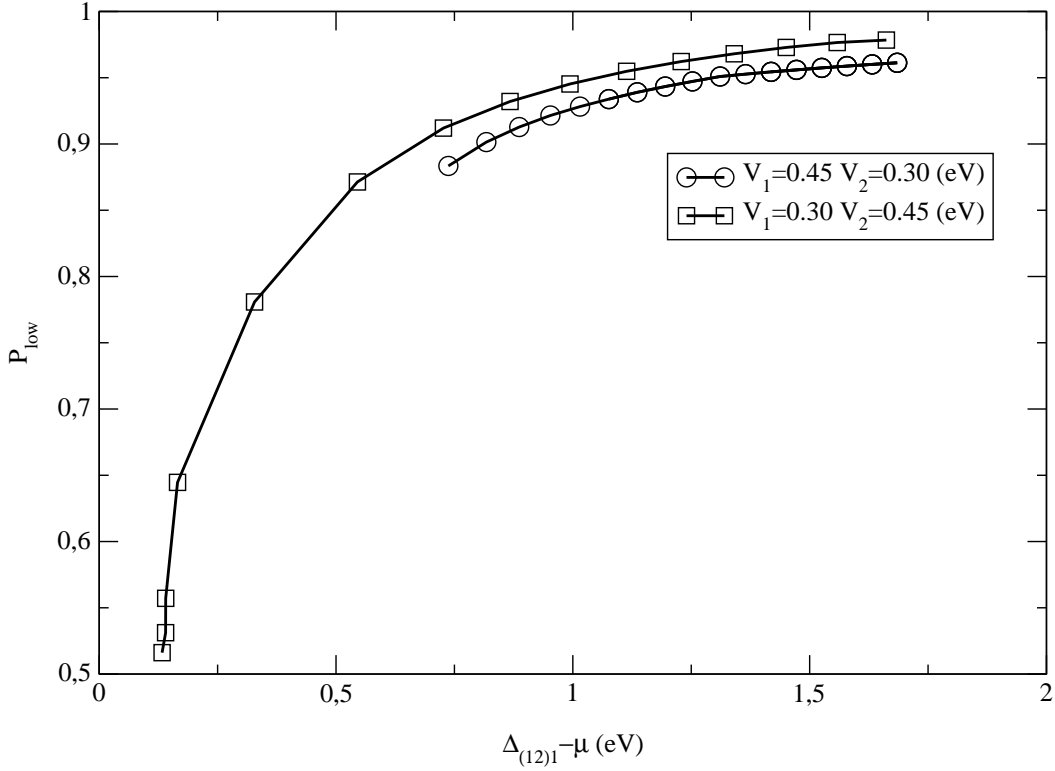}
\caption{Population in lower pole, when changing the parameters
$\Delta_{(12)1}-\mu$.}
\label{p_D-mu:fig}
\end{figure}
\end{center}

For the constant DOS we choose the bandwidth to be 8 eV.
The population of the lower level for different cases of the mixing 
is shown in Fig.\ref{pv1pv2:fig}.
From this figure we see that the 
population drops when the mixing is increased, but, there is also an 
interesting 
feature for $V_2\simeq 0.2$ eV and above. When increasing $V_1$, 
there is actually a 
{\it localization} of the electrons. Since $V_2$ comes from the bandstructure 
($V_1$ is the local mixing coming from the Coulomb 
interaction) we see that changes in 
structure or composition which reduce $V_2$, can actually mean that 
we have a larger degree of delocalization. Hence, delocalization does not 
always 
come from {\it increased} mixing interaction, but decreased.
The Hubbard $U$-parameter was extracted and the result is shown in 
\begin{center}
\begin{figure}[hbt]
\includegraphics[angle=270,scale=0.6]{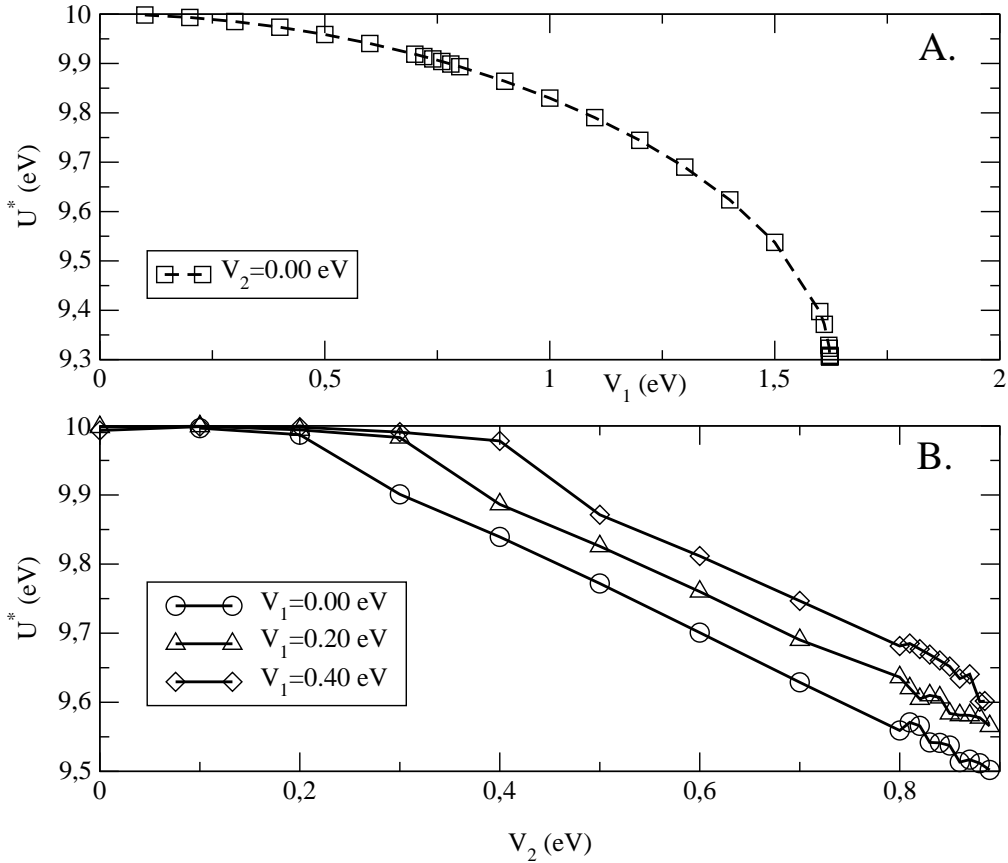}
\caption{Effective Hubbard $U$, $U^*$, when changing the parameters $V_1$ and
$V_2$. }
\label{U:fig}
\end{figure}
\end{center}
Fig.\ref{U:fig}A and B, where $V_1$, and $V_2$ is changed respectively. 
We see that $U$ decrease with almost 10\% 
when mixing is changed. This actually moves the upper transition closer to 
the Fermi-level, and therefore also affects the delocalization.
There is, however, another important effect, the distance between the upper 
transition, $\Delta_{(12)1}$, and the chemical potential $\mu$. 
The effect of this is shown in Fig.\ref{p_D-mu:fig}.
In this figure, we see that one driving mechanism of delocalization 
is the parameter $\Delta_{(12)1}-\mu$. Further, we see that the behavior is 
almost independent for the different values of mixing. The best fit to 
the $P_{low}$ ($P_{low}\stackrel{{\rm def}}{=}N_1+N_2$) 
versus $\Delta_{(12)1}-\mu$ is obtained from the function
\begin{equation}
P_{low}=\alpha+\frac{\beta}{(\Delta_{(12)1}-\mu)+\gamma}.
\label{Plow:eq}
\end{equation}
Where $\alpha, \beta$ and $\gamma$ are constants with the values given in 
table~\ref{abg:table}.
\begin{table}[hbt]
\begin{tabular}{ccc}
$\alpha$ & $\beta$  & $\gamma$ \\ \hline
1.03465 & -0.0949886 &  0.0551839 \\
\end{tabular}
\caption{Values of the constants determining $P_{{\rm low}}$, in
Eqn.~\ref{Plow:eq}.}
\label{abg:table}
\end{table}
Using this equation we determine the value of $\Delta_{(12)1}-\mu$ when 
complete delocalization occurs to be 0.0366 eV. i.e.\ when the transition is 
very close to the chemical potential. 
\subsection{LMTO-DOS Case}
\begin{figure}[hbt]
\includegraphics[scale=1.0]{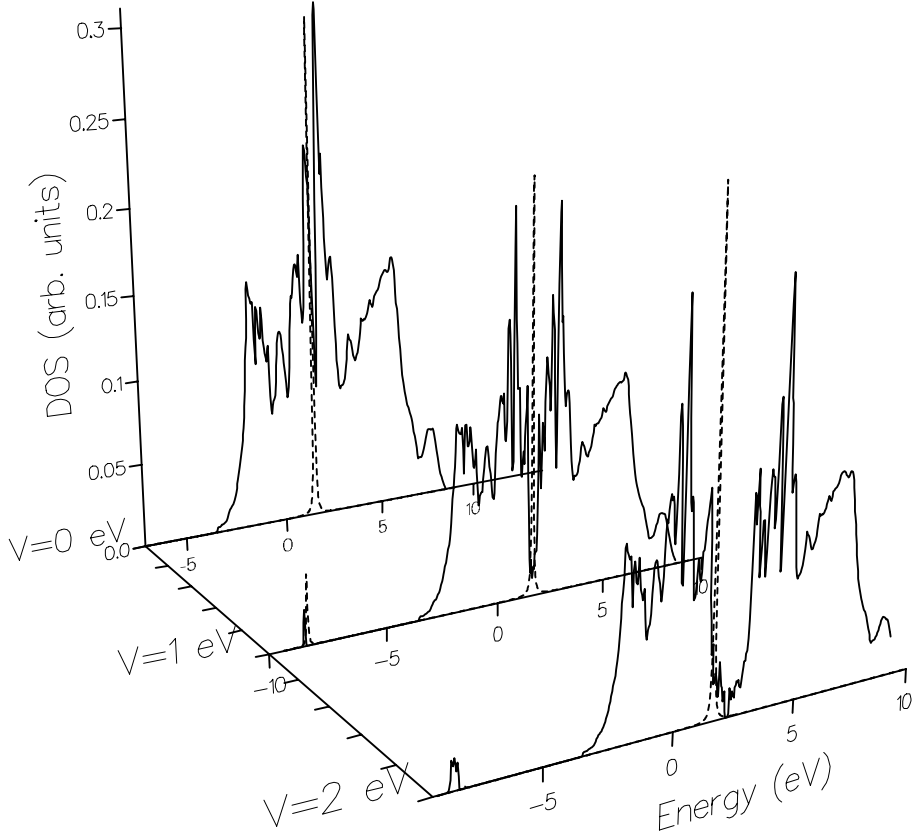}
\caption{
The delocalization and changes in DOS for different values of the mixing
parameter. ($V_2=0$, and $V=V_1$ in the figure). }
\label{result2}
\end{figure}

When we used the DOS from a band-structure calculation we obtained the 
following result. For this case we put $V_2=0$. 
The DOS after the self-consistent cycle is shown in
Fig.\ref{result2}. As seen, 
upon an increase of mixing, the  
bandwidth is narrowed and a gap where the upper level is situated is
opened up. 
The states, pushed out of the gap region, enhance the DOS at
the Fermi level. The gap that develops and the delocalization is further
illustrated in Fig.\ref{result2}. 
The peak above the Fermi level will not be seen in experiments since these 
transitions become delocalized due to mixing interactions.  
\begin{center}
\begin{figure}[hbt]
\includegraphics[angle=270,scale=0.6]{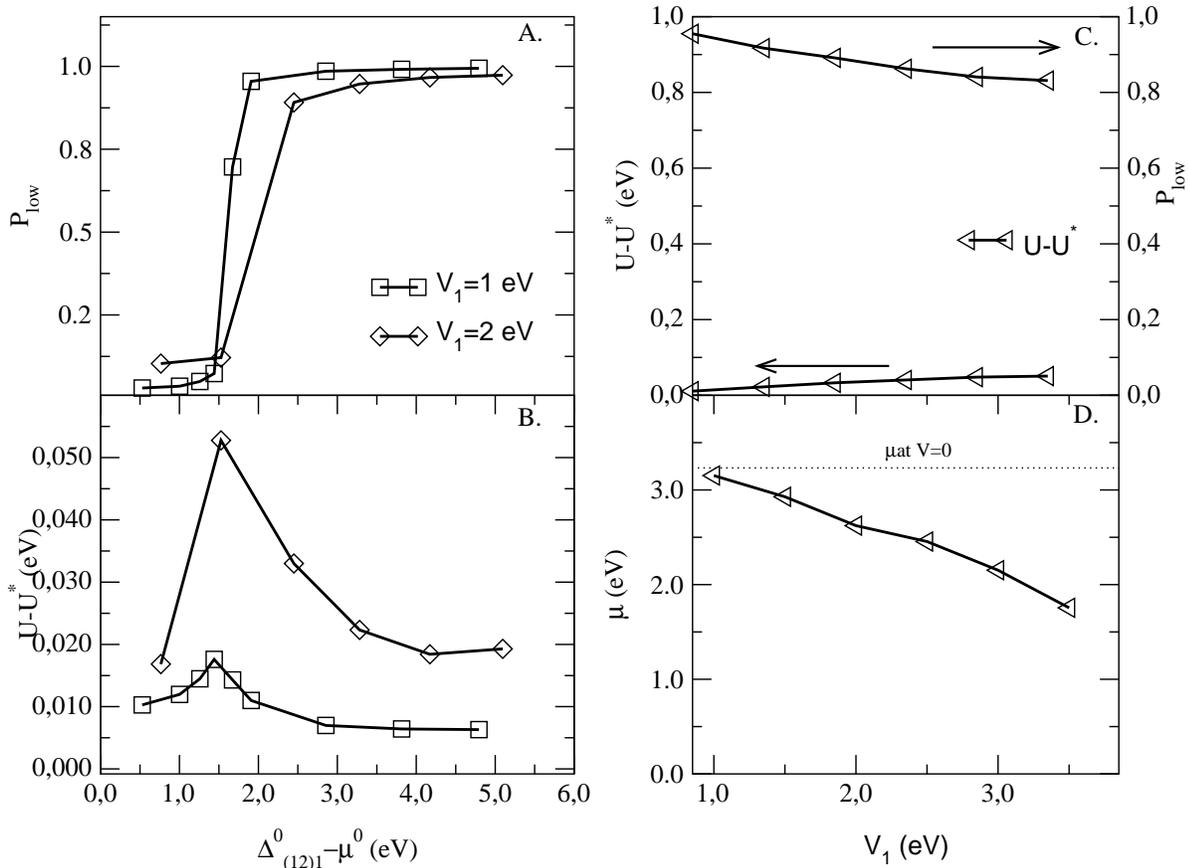}
\caption{Changes in $N_{(12)}$, $U$ and $\mu$ for different choices of
parameters.  ($V_2=0$).}
\label{result}
\end{figure}
\end{center}
%
In Fig.\ref{result} we present the result of the self-consistent 
calculations. As seen from equations (\ref{freq:Pr}) and
(\ref{pop:Pr}), the energy difference between the upper level and the chemical
potential is one of the most crucial parameters of the model. 
Fig.\ref{result}a shows the dependence of the population number $N_{12}$ 
on this parameter for two values of mixing. The Hubbard parameter
chosen is 10 eV and the mixing, $V_{\bf k}=V_1$, is 1 and 2 eV
respectively; the bandwidth is 9 eV.
The next figure, Fig.\ref{result}b, shows 
how $U$, defined as $\Delta_{(12\nu)12}-\Delta_{(12)1}$, 
change as a function of the same parameters. The graph in 
Fig.\ref{result}d shows the 
change of the chemical potential $\mu$ and the fourth
graph, Fig.\ref{result}c, shows the change in $U$ and population 
number $N_{12}$ when only mixing is changed. 
As seen, in Fig.\ref{result}c, dependences on the value of the 
mixing parameter are smooth and the
only instability is caused by the level slightly above the Fermi energy. 
The value of the critical parameter $\Delta_{(12)(12\nu)}^0-\mu$ can be
changed by chemical doping, which will shift the chemical potential and
therefore affect the transition pressure. 
If we look at the form of the DOS, generated by pure LDA calculations 
(see Fig.\ref{result2} for V=0), it does
not contain a quasi-gap in the vicinity of $\epsilon_f$, since within 
band structure calculations all \f-states are taken in core and the 
transition $\Delta_2$ does not exist. It is, however, worth to note that 
experimentally photo-electron spectroscopy shows peaks above the Fermi level 
for all rare earths. These peaks are interpreted as \f-states. If we 
treat the \f-states as fully delocalized we have peaks above the 
Fermi-level but 
the equilibrium properties are not well described (see e.\ g.\ 
ref.~\cite{lundin}).   

\section{Conclusion}
We have studied the scenario of \f-electron delocalization and 
dependence of Hubbard $U$ on the value of the mixing interaction which is
assumed to increase proportionally to the external 
pressure. The scenario considered emphasizes that the dominant role 
is played by the \f-sum rule, connecting
excitations in different energy regions and resonant
mixing of ($spd$)-electrons and Hubbard's \f-excitations. The mechanism can
only exist due to strong correlations. Thus, in our picture, 
delocalization arises due to the following reasons: 
First, the strong correlations provide the separation of the \f-shell into two 
manifolds in different energy regions. Second, under the application of 
pressure the mixing increases due to an increase in the overlap. 
Third, the shift of bands, and the renormalization of 
the levels, alter the crucial parameter $\Delta_{(12)1}-\mu$.  

\section*{Acknowledgments}
This project has been financed by the Swedish Natural Science Research 
Council (NFR), and Materials Consortium No.\ 9.\ I.\ S.\ thanks for 
support from the G\"oran Gustafsson foundation.

\end{document}